\documentclass[11pt,a4paper]{article}

\usepackage[utf8]{inputenc}
\usepackage[T1]{fontenc}
\usepackage{amsmath,amssymb}
\usepackage{graphicx}
\usepackage{booktabs}
\usepackage{hyperref}
\usepackage{url}
\usepackage{natbib}
\usepackage{authblk}
\usepackage[margin=2.5cm]{geometry}
\usepackage{caption}
\usepackage{subcaption}
\usepackage{float}
\usepackage{enumitem}
\usepackage{xcolor}
\usepackage{tikz}
\usetikzlibrary{positioning,arrows.meta,calc,fit,backgrounds,shapes.geometric}

\hypersetup{
  colorlinks=true,
  linkcolor=blue,
  citecolor=blue,
  urlcolor=blue
}

\tikzset{
  layer/.style={draw, rounded corners, thick, align=center, inner sep=6pt},
  daskbox/.style={layer, fill=blue!6, draw=blue!55},
  extbox/.style={layer, fill=orange!8, draw=orange!60},
  llmbox/.style={layer, fill=green!8, draw=green!55},
  databox/.style={layer, fill=gray!10, draw=gray!55},
  flow/.style={-{Stealth[length=2.4mm]}, thick},
  flowdash/.style={-{Stealth[length=2.4mm]}, thick, dashed},
}

\title{\textbf{Smallholder Farmers on Remote Islands Receiving Retrieval-Grounded Multilingual LLM Assistance and Agronomic Advice}}

\author[1]{Nikolaos~D.~Tantaroudas\thanks{Corresponding author: \texttt{nikolaos.tantaroudas@iccs.gr}}}
\author[2]{Ilias~Karachalios}
\author[3]{Andrew~J.~McCracken}

\affil[1]{Institute of Communications and Computer Systems (ICCS), Iroon Polytechneiou 9, 15773 Zografou, Athens, Greece}
\affil[2]{National Technical University of Athens, Zografou, 15772 Athens, Greece}
\affil[3]{DASKALOS-APPS, 183 Rue de l'Abb\'{e} Griffon, 01960 P\'{e}ronnas, France}

\date{}

\begin{document}

\maketitle

\begin{abstract}
Smallholder farming communities in remote and depopulating areas have limited access to agricultural advisory services, and their locally specific agronomic knowledge, often expressed in regional dialect, is poorly represented in the global corpora on which modern Large Language Models (LLMs) are trained. A general-purpose chatbot consequently answers questions about such a place fluently but unreliably, with no grounding in authoritative local data that a farmer can trust. This paper presents a conversational AI assistant, ``Falco eleonorae'', embedded within a bilingual (Greek-primary, English-secondary) e-market platform that serves the farmers and cooperatives of a defined island area of interest. The assistant is engineered as a thin Backend-for-Frontend (BFF) proxy in front of a geospatially-aware agronomic agent rather than as a self-hosted model. Answer generation and tool selection are delegated to a managed upstream service built on OpenAI GPT-5-family models, while a single bounded task, producing a short textual description of an uploaded field photograph, is handled directly by a vision-capable model (Anthropic Claude Sonnet 4.5) so that only text ever reaches the agent, and voice input is transcribed by a managed streaming speech-to-text service hosted in an EU region. The assistant is grounded not through a self-hosted vector database but through tool-augmented retrieval, in which a Model Context Protocol (MCP) tool queries a curated, read-only, bilingual data interface exposing local crops, a seasonal calendar, traditional practices, a dialect glossary, products, agritourism experiences, cooperatives, and training content, and every query is wrapped in a geospatial Well-Known Text envelope that anchors the agent to the area of interest. We detail the conversational interface together with its multilingual, voice, and image modalities, its progressive-web-application and accessibility design for low-bandwidth field use, and its security and data-protection posture, and we argue that for a small and resource-constrained rural deployment a managed and grounded multilingual assistant of this kind is both more attainable and more trustworthy than a bespoke self-hosted model.
\end{abstract}


\section{Introduction: The Challenge}
\label{sec:intro}

Kythera and its smaller neighbour Antikythera are remote islands at the southern edge of the Aegean. Like many European island and mountain regions, they combine a depopulating and ageing agricultural workforce, a fragmented set of small producers, seasonal agritourism, and a body of locally specific agronomic and cultural knowledge, including a distinctive regional dialect, that is largely absent from mainstream digital resources. The institutional channel that would ordinarily carry agronomic knowledge to such producers, the Agricultural Knowledge and Innovation System (AKIS), is, in thinly populated regions, correspondingly thin, with advisory capacity that is scarce, expensive to deliver in person, and difficult to sustain~\citep{akis2024review}. The 2023--2027 Common Agricultural Policy explicitly elevates AKIS and the digitalisation of advisory services as a cross-cutting objective, precisely because this gap is structural rather than incidental.

Large Language Models (LLMs) are an obvious candidate technology for narrowing this gap, and recent work has begun to assess their use in agricultural extension~\citep{tzachor2023llm,high2026aiextension}. An LLM can render scientific guidance in plain language, answer follow-up questions, operate in the user's own language, and be always available. Yet a naive deployment of a general-purpose chatbot is poorly matched to the island smallholder setting for several reasons. The most fundamental is place-agnostic knowledge and the attendant risk of hallucination, since a general LLM has at best sparse and possibly outdated knowledge of an island's specific crop varieties, harvest windows, traditional practices, local producers, and dialect terms, so that when asked a local question it tends to answer fluently but unreliably, with no grounding in authoritative local data that the farmer can trust~\citep{tzachor2023llm}. Language is a second concern, because the community is Greek-speaking, with Greek as the primary language of interaction and English a secondary one for visitors and external partners, so advisory tooling must be Greek-first. Digital literacy and access form a third, as many prospective users have limited digital experience and intermittent connectivity, which requires an interface that tolerates poor networks, is usable on a phone in the field, and supports modalities, voice in particular, that lower the barrier for users uncomfortable with typing. Finally, trust and governance are paramount, because for an EU-funded public-interest service handling the personal data of a small, identifiable community, data residency, privacy, and the avoidance of credential or content leakage are first-order requirements rather than optional hardening.

This paper describes how such an assistant addresses these constraints. It is embedded in a bilingual platform that is delivered for, and grounded in the local knowledge of, the island of Kythera, combining an e-market for local products and agritourism experiences with a training and certification system, a knowledge repository, and an analytics layer, built for the Kythera pilot under the European \mbox{PoliRuralPlus} initiative~\citep{poliruralplus}. Our focus is the AI assistant itself, ``Falco eleonorae'' (after the Eleonora's falcon that breeds on the islands), the user-facing front end to a geospatially-aware agronomic agent operated within the \mbox{PoliRuralPlus} ecosystem and known as ``Jackdaw''.

Rather than train or self-host a model, the assistant is deliberately architected as a thin Backend-for-Frontend (BFF) proxy in front of a managed agent, and grounds that agent in local knowledge through a structured data interface.  The paper describes an implemented and deployed LLM-assistance architecture for a rural pilot in which no model is self-hosted, with the answer and tool-selection LLMs running upstream as a managed service and the platform calling a frontier model directly only for a single bounded multimodal task (Section~\ref{sec:architecture}). It then gives a concrete account of tool-augmented retrieval as an alternative to classic vector-store retrieval-augmented generation (RAG) for grounding an agent in small, curated, frequently-changing local knowledge, realised through a Model Context Protocol (MCP) tool over a read-only bilingual data API and wrapped in a geospatial area-of-interest envelope (Section~\ref{sec:grounding}). It finally describes the conversational interface and its multilingual, voice, and image modalities, designed for low digital literacy and poor connectivity, together with the system's security and data-protection design (Sections~\ref{sec:interface}--\ref{sec:impl}). 

\section{Background and Related Work}
\label{sec:related}

\subsection{Large language models}
\label{sec:bg_llm}

Modern conversational LLMs descend from the Transformer architecture~\citep{vaswani2017attention}, whose self-attention mechanism replaced recurrence and enabled the large-scale, parallelisable pre-training of decoder-only language models. Scaling such models with autoregressive pre-training yielded strong few-shot ``in-context'' learning~\citep{brown2020language}, and aligning them to follow instructions via reinforcement learning from human feedback (RLHF) produced the assistant-style behaviour now familiar from production systems~\citep{ouyang2022instructgpt}. A complementary alignment approach, Constitutional AI, derives harmlessness from AI feedback against an explicit written set of principles rather than solely from human preference labels~\citep{bai2022constitutional}. The two model families relevant to this work, OpenAI's GPT-5 family and Anthropic's Claude, are the products of this lineage; their exact parameter counts are not publicly disclosed by their vendors. These models are treated here as managed capabilities described by their public model and system cards~\citep{openai2025gpt5card,anthropic2025sonnet45}.

\subsection{Grounding LLMs: retrieval and tool use}
\label{sec:bg_rag}

Because a parametric language model's knowledge is fixed at training time and prone to confabulation, grounding it in an external knowledge source is essential for factual, domain-specific assistance. The canonical mechanism is retrieval-augmented generation (RAG)~\citep{lewis2020rag}, which couples a parametric generator with a non-parametric retriever, typically dense passage retrieval over an embedded document corpus~\citep{karpukhin2020dpr}. RAG excels when the knowledge base is a large, relatively static text collection. An alternative, increasingly favoured for structured or rapidly-changing data, is \emph{tool-augmented} retrieval, in which the model interleaves reasoning with calls to external tools or APIs and incorporates their results into its answer; the ReAct paradigm formalised this reasoning-and-acting loop~\citep{yao2023react}. The Model Context Protocol (MCP) standardises how such tools are exposed to an LLM application over a JSON-RPC interface, decoupling the tool provider from the model host~\citep{mcp2025spec}. A related strand of work has the model emit executable queries against structured data directly, rather than retrieving unstructured passages~\citep{tantaroudas2026querying}, which is precisely the regime in which the grounding tool described here operates. As we argue in Section~\ref{sec:grounding}, the assistant uses tool-augmented retrieval over a curated structured API, not vector-store RAG, and we motivate that choice by the nature of the knowledge being grounded.

\subsection{AI for agricultural extension}
\label{sec:bg_ag}

Tzachor et~al.~\citep{tzachor2023llm} assessed GPT-based models for agricultural extension, finding genuine promise in simplifying scientific guidance and personalising location-specific advice, but stressing the necessity of human experts in the loop and of guarding against confidently wrong, context-free output, exactly the failure mode that grounding is meant to prevent. High et~al.~\citep{high2026aiextension} analysed a deployed farmer-facing chatbot across several countries and found that, while such tools improve access and personalisation through multimodal interfaces, their effectiveness depends on embedding them within trusted social infrastructure and aligning them with participatory extension principles rather than one-way information delivery. These findings directly inform our design, namely to ground the model in curated local data, keep the content under the editorial control of the local pilot organisation, and prioritise accessible multimodal interaction. The present authors have previously built and validated integrated AI service platforms that compose managed AI capabilities behind a thin, security-conscious application layer, spanning immersive accessibility and communication~\citep{tantaroudas2026interact}, personalised career guidance~\citep{tantaroudas2026career}, and AI-driven cybersecurity~\citep{tantaroudas2026sentinel}; the current work applies the same engineering philosophy to rural agricultural knowledge.

\section{System Architecture}
\label{sec:architecture}

\subsection{A four-layer, no-self-hosted-model design}
\label{sec:arch_overview}

\begin{figure}[htbp]
\centering
\begin{tikzpicture}[font=\footnotesize]
\node[daskbox, text width=8.4cm] (dask) at (0,0) {\textbf{Layer 1 -- e-market platform (Django BFF; no chat LLM)}\\[2pt]
  \footnotesize Chat UI (PWA) \ \textbullet\ \ session + history \ \textbullet\ \ image description \ \textbullet\ \ voice-to-text \ \textbullet\ \ SSE proxy};
\node[extbox, text width=8.4cm] (bff) at (0,-3.0) {\textbf{Layer 2 -- \texttt{jackdaw.online} (agent gateway / BFF)}\\[1pt]\footnotesize SSE chat endpoint; injects the user's upstream session};
\node[extbox, text width=8.4cm] (api) at (0,-5.6) {\textbf{Layer 3 -- \texttt{prp-api} (agent backend; owns the LLMs)}\\[2pt]
  \footnotesize \textbf{Answer LLM:} OpenAI \texttt{gpt-5-nano} \ \textbullet\ \ \textbf{Tool selector:} OpenAI \texttt{gpt-5-mini}\\
  \footnotesize multi-step reasoning loop; WKT (SRID~4326) geo-envelope on every request};
\node[extbox, text width=5.0cm, anchor=west] (raven) at (-4.2,-8.3) {\textbf{Layer 4 -- \texttt{prp-raven}}\\[1pt]\footnotesize MCP tool server: the \texttt{kythera} tool (9 sub-tools)};
\node[llmbox, text width=4.7cm, anchor=west] (claude) at (4.9,-1.0) {\textbf{Anthropic Claude Sonnet 4.5}\\[1pt]\footnotesize image description (called \emph{directly} by the platform)};
\node[llmbox, text width=4.7cm, anchor=west] (aws) at (4.9,-2.7) {\textbf{Amazon Transcribe (EU)}\\[1pt]\footnotesize streaming voice-to-text};
\node[databox, text width=3.4cm, anchor=west] (kapi) at (4.9,-8.3) {\textbf{Platform \texttt{/api/v1/}}\\[1pt]\footnotesize read-only bilingual local data};
\draw[flow] (dask) -- node[left]{chat turn (SSE)} (bff);
\draw[flow] (dask.east) -- node[above,sloped,font=\scriptsize]{image} (claude.west);
\draw[flow] (dask.east) -- node[below,sloped,font=\scriptsize]{voice} (aws.west);
\draw[flow] (bff) -- (api);
\draw[flow] (api.south) -- node[left,pos=0.6]{MCP tool calls} (raven.north);
\draw[flow] (raven.east) -- node[above,font=\scriptsize]{HTTP+key} (kapi.west);
\draw[flowdash] (kapi.east) -- (8.9,-8.3) -- node[right,pos=0.55]{grounding data} (8.9,0) -- (dask.east);
\end{tikzpicture}
\caption{The four-layer architecture. The platform (layer~1) hosts no chat LLM; it proxies the conversational turn to the upstream \mbox{PoliRuralPlus} agent (layers~2--3), which runs the OpenAI GPT-5-family models and orchestrates tool use, and grounds answers via an MCP tool (layer~4) that queries the platform's own read-only local data API. The platform calls Anthropic Claude Sonnet~4.5 directly only for image description, and Amazon Transcribe for voice.}
\label{fig:architecture}
\end{figure}
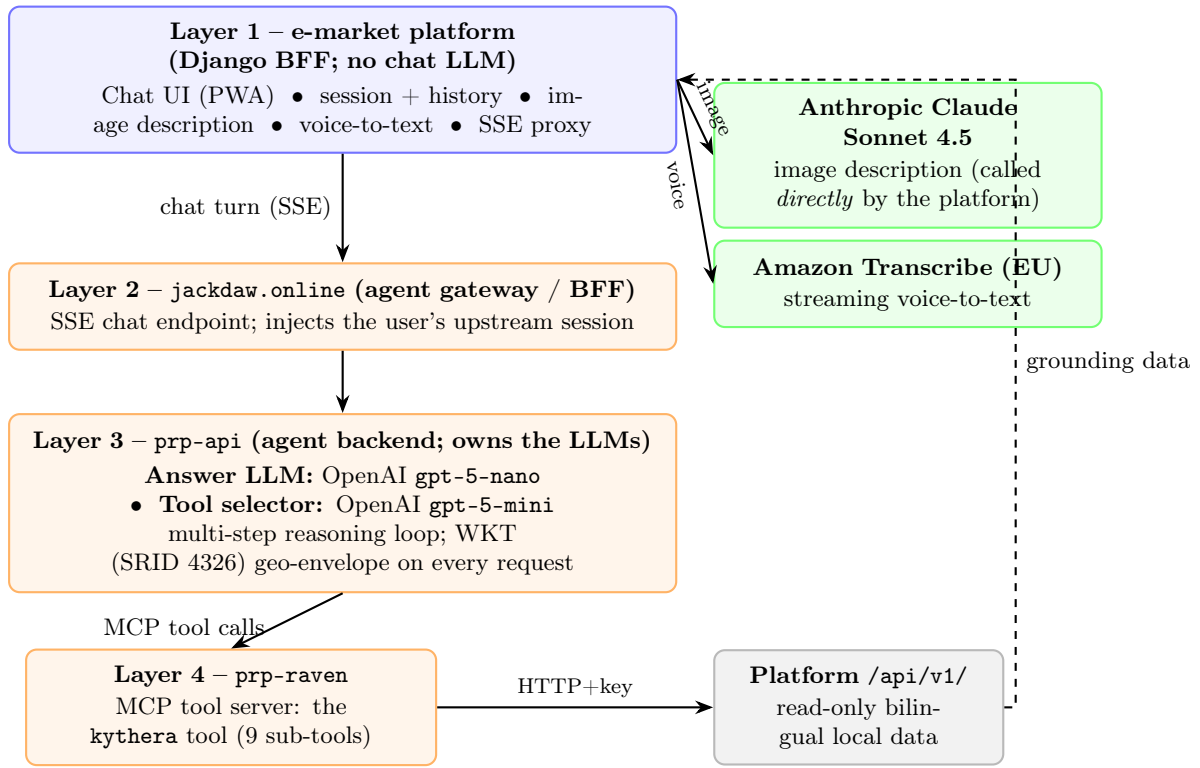

The assistant is organised into four layers (Figure~\ref{fig:architecture}). The platform itself (layer~1) is a Django application that acts as a Backend-for-Frontend that renders the chat interface, manages the user session and conversation history, performs two bounded multimodal pre-processing tasks (image description and speech-to-text), and proxies the conversational turn to the upstream agent. It runs \emph{no} chat LLM of its own. Layers 2--4 are external services operated within the \mbox{PoliRuralPlus} ecosystem: a BFF/agent gateway (\texttt{jackdaw.online}); the agent backend itself (\texttt{prp-api}), a service that hosts the LLMs, performs tool selection, and orchestrates the multi-step reasoning loop; and a catalogue/tool server (\texttt{prp-raven}) that exposes domain tools over MCP. Crucially, one of those MCP tools, contributed by the platform team, reaches back into the platform's own read-only data API to retrieve area-specific knowledge (Section~\ref{sec:grounding}).

\subsection{The model stack}
\label{sec:arch_models}

Table~\ref{tab:models} enumerates the models and where each runs. Three points are worth emphasising for accuracy. First, the conversational ``brain'' (answer generation and tool selection) is \emph{not} hosted by the platform; it is the upstream agent's responsibility, and the upstream service owns the associated billing and rate limits. Second, the only frontier model the platform invokes directly is Anthropic Claude Sonnet~4.5, used for a single, bounded task, converting an uploaded field photograph into a short factual textual description that is then prepended to the user's next text turn. Third, voice input is transcribed by a managed streaming service (Amazon Transcribe) pinned to an EU region, and no self-hosted speech model is run. 

\begin{table}[htbp]
\centering
\caption{The model stack. ``Where'' indicates which party runs the model. Parameter counts for all listed proprietary models are undisclosed by their vendors.}
\label{tab:models}
\small
\begin{tabular}{p{3.5cm} p{4.6cm} p{4.6cm}}
\toprule
\textbf{Model / service} & \textbf{Role} & \textbf{Where it runs} \\
\midrule
OpenAI \texttt{gpt-5-nano} & Primary chat / answer-text generation & Upstream (\texttt{prp-api}); not hosted by the platform \\
OpenAI \texttt{gpt-5-mini} & Tool selection / routing in the agent loop & Upstream (\texttt{prp-api}) \\
Anthropic Claude Sonnet~4.5 & Image description (photo $\rightarrow$ short text note) & Platform-side, in-process, via the vendor SDK \\
Amazon Transcribe (streaming) & Voice-to-text for the chat microphone & Platform-side / AWS, EU region \\
\bottomrule
\end{tabular}
\end{table}

\subsection{Why a thin proxy?}
\label{sec:arch_rationale}

The decision not to self-host a model is a deliberate fit to the pilot's constraints. A small island pilot has neither the inference budget nor the operational staff to run and maintain a frontier model; a managed upstream agent already provides the geospatial agronomic tools (land cover, climatology, terrain, demography) that a GeoAI assistant needs; and concentrating the platform's own direct model use in a single, well-scoped task keeps the attack surface, the cost surface, and the data-egress surface small and auditable. The cost of this choice is a hard dependency on external services, which we treat candidly as a limitation (Section~\ref{sec:limits}).

\subsection{The geospatial envelope}
\label{sec:arch_wkt}

The upstream agent is a GeoAI system: it expects every request to carry a Well-Known Text (WKT) geometry (SRID~4326) describing the area of interest, and rejects requests without one. The platform proxy therefore attaches a geometry to every turn. When the user has not selected a map area, the platform supplies a default polygon corresponding to the Kythera/Antikythera bounding box, ensuring that even a non-geospatial question (``which olive varieties grow here?'') is answered in the correct geographic context rather than for an arbitrary location. Spatially-aware tools (Section~\ref{sec:grounding}) use this envelope to restrict and rank results to the relevant area.

\section{Grounding the Model in Local Knowledge}
\label{sec:grounding}

\subsection{Tool-augmented retrieval, not a vector store}
\label{sec:grounding_choice}

A central design question for any knowledge assistant is how the model is grounded. The default answer in much current practice is vector-store RAG~\citep{lewis2020rag,karpukhin2020dpr}, which embeds a document corpus, retrieves the nearest passages to the query, and conditions generation on them. The platform deliberately does \emph{not} do this. There is no embedding pipeline and no vector database in it. Instead, grounding is realised as \emph{tool-augmented retrieval}~\citep{yao2023react} over a curated, structured, read-only data interface, mediated by MCP~\citep{mcp2025spec}.

The motivation is the nature of the knowledge to be grounded. The island's authoritative knowledge is not a large static text corpus; it is a comparatively small, curated, frequently-edited, and inherently structured body of records, namely crops and their seasons, traditional practices, dialect glossary entries, local products, agritourism experiences, cooperatives, and training lessons, maintained by the pilot's editors. For knowledge of this shape, exact structured queries (the seasonal-calendar entries for a given month) are more faithful and more current than approximate nearest-neighbour search over embeddings~\citep{tantaroudas2026querying}, and they inherit, for free, the editorial controls that already govern the records, so that drafts are excluded, hidden prices are not revealed, and a content edit is reflected on the very next query with no re-indexing. The agent's general agronomic reasoning is still supplied by the upstream LLM; what the structured tool guarantees is that, when the question is local, the answer is anchored in the pilot's own authoritative data.

\subsection{The local data API and MCP tool}
\label{sec:grounding_api}

The platform exposes a read-only, bilingual JSON data API (Table~\ref{tab:endpoints}). A dedicated MCP tool, the \texttt{kythera} tool, contributed by the platform team to the upstream catalogue server, presents this API to the agent as a family of nine callable sub-tools (knowledge search, crop information, seasonal calendar, glossary lookup, product search, experience search, cooperative search, training-module listing, and lesson-content retrieval). When the upstream tool-selector model judges a turn to be locally scoped, it invokes the appropriate sub-tool; the tool calls the platform's data API over HTTP with a shared secret key; and the structured result is returned into the agent's reasoning loop. Two of the sub-tools (experiences and cooperatives) are spatially aware, parsing the WKT envelope, restricting results to that bounding box, and ordering them by proximity to its centroid.

\begin{table}[htbp]
\centering
\caption{The read-only bilingual local data API surfaced to the agent through the \texttt{kythera} MCP tool. Every endpoint returns Greek or English content (Greek fallback) and enforces editorial gating (drafts and inactive records excluded; hidden prices suppressed).}
\label{tab:endpoints}
\small
\begin{tabular}{p{5.0cm} p{8.5cm}}
\toprule
\textbf{Endpoint} & \textbf{Local knowledge exposed} \\
\midrule
\texttt{knowledge/search} & published knowledge articles, traditional practices, FAQs \\
\texttt{knowledge/seasonal-calendar} & seasonal calendar entries by month (optionally by crop) \\
\texttt{crops/\{type\}} & crop record plus related practices and articles \\
\texttt{glossary/lookup} & glossary entries, including regional dialect terms \\
\texttt{products/search} & local catalogue products (public) \\
\texttt{experiences/search} & agritourism experiences (spatially aware) \\
\texttt{cooperatives/search} & cooperative directory (spatially aware) \\
\texttt{training/modules}, \texttt{.../lessons/\{slug\}} & training-module listings and full lesson content \\
\bottomrule
\end{tabular}
\end{table}

\subsection{Tool selection and query formation}
\label{sec:grounding_selection}

A natural question is how the agent decides which tool to call and how it obtains the search terms that the tool needs. The mechanism is two LLM-driven steps rather than an explicit keyword-extraction pipeline. First, tool selection is performed by \emph{semantic routing}: a dedicated tool-selector model (OpenAI \texttt{gpt-5-mini}, run at minimal reasoning effort) is presented with the user's natural-language message together with the catalogue of available tools, each described only by a name and a short natural-language description, and it selects the one or two tools whose descriptions best match the \emph{intent} of the question. There is no separate stage that first extracts keywords and then looks them up; the routing is by meaning, which is why the quality of each tool's description is decisive. A tool is chosen for a question about harvesting olives because its description advertises crop and seasonal-calendar knowledge, not because the token ``olive'' was matched against an index. The locally-scoped system prompt assembled by the platform (Section~\ref{sec:grounding_turn}) steers this routing by naming the local \texttt{kythera} sub-tools and instructing the agent to prefer them over its generic dataset searches for any local question, while capping the number of tool calls at two per turn to bound latency and cost.

Second, once a tool is chosen, the agent populates that tool's typed parameters from the question, and it is at this point that the search terms, the ``keywords'', are actually formed. For a free-text sub-tool such as \texttt{search\_kythera\_knowledge(query, lang, limit)} the model writes the salient terms of the question into the \texttt{query} argument; for a structured sub-tool such as \texttt{get\_seasonal\_calendar(month, crop\_type, lang)} it fills the discrete filters, for example a month and a crop, inferred from the question; and the response language is taken from the active interface language carried in the system prompt. The geospatial parameters are not produced by the model at all but injected automatically from the active map area-of-interest as the Well-Known Text envelope of Section~\ref{sec:arch_wkt}. The chosen tool then calls the read-only data API, which performs the actual matching, substring matching on the free-text \texttt{query} and exact matching on the structured filters, and returns the records that match. Those records, rather than the model's parametric memory, are what the answer model then uses to compose its reply, surfacing any source link the tool returned. This division of labour, semantic routing to choose a tool and parameter synthesis to query it, is an instance of the tool-augmented retrieval pattern of \citet{yao2023react}, and the use of a language model to emit an executable structured query rather than to retrieve unstructured passages mirrors recent work on natural-language querying of structured data~\citep{tantaroudas2026querying}.

\subsection{Anatomy of a grounded turn}
\label{sec:grounding_turn}

\begin{figure}[htbp]
\centering
\begin{tikzpicture}[font=\footnotesize]
\node[daskbox] (b) at (0,0) {Browser};
\node[daskbox] (p) at (3.2,0) {Platform proxy};
\node[extbox] (g) at (6.6,0) {Gateway};
\node[extbox] (a) at (9.4,0) {Agent (LLMs)};
\node[databox] (k) at (12.6,0) {Local API};
\foreach \n in {b,p,g,a,k}{\draw[gray!50,dashed] (\n.south) -- ++(0,-5.2);}
\begin{scope}[every path/.style={flow}]
\draw ($(b.south)+(0,-0.4)$) -- node[above]{1. message} ($(p.south)+(0,-0.4)$);
\draw ($(p.south)+(0,-1.0)$) -- node[above]{2. prompt + history + WKT} ($(g.south)+(0,-1.0)$);
\draw ($(g.south)+(0,-1.6)$) -- node[above]{3. forward} ($(a.south)+(0,-1.6)$);
\draw ($(a.south)+(0,-2.2)$) -- node[above]{4. tool call} ($(k.south)+(0,-2.2)$);
\draw ($(k.south)+(0,-2.8)$) -- node[below]{5. records} ($(a.south)+(0,-2.8)$);
\draw ($(a.south)+(0,-3.4)$) -- node[below]{6. grounded answer (SSE)} ($(g.south)+(0,-3.4)$);
\draw ($(g.south)+(0,-4.0)$) -- node[below]{7. relay} ($(p.south)+(0,-4.0)$);
\draw ($(p.south)+(0,-4.6)$) -- node[below]{8. stream + persist} ($(b.south)+(0,-4.6)$);
\end{scope}
\end{tikzpicture}
\caption{Sequence of a single grounded conversational turn. Tool selection and answer generation occur upstream (steps 3--6); the platform proxy contributes the locally-scoped system prompt and geo-envelope (step~2) and the local grounding data (steps 4--5), and relays the streamed answer (steps 7--8).}
\label{fig:sequence}
\end{figure}
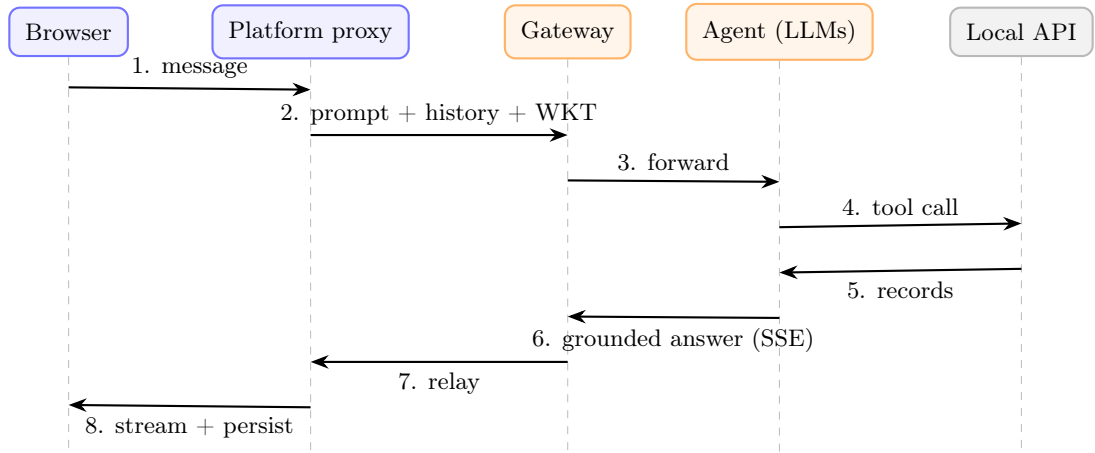

Figure~\ref{fig:sequence} traces a single grounded turn. The browser sends the user's message to the platform proxy, which assembles a request comprising a freshly-built, locally-scoped system prompt (encoding the user's profile and language, and instructing the agent to prefer the local \texttt{kythera} tools over generic dataset searches), the recent conversation history, and the WKT geo-envelope. The proxy opens a streaming connection to the upstream gateway. Upstream, the tool-selector model chooses tools; the \texttt{kythera} MCP tool calls back into the platform's data API; the answer model composes a reply grounded in the returned records; and the gateway streams the result back as Server-Sent Events (SSE), which the proxy relays verbatim to the browser. The assistant turn, and a cross-turn thread identifier that lets the upstream agent retain conversational state, are persisted by the platform.

\section{The Conversational Interface}
\label{sec:interface}

\begin{figure}[htbp]
\centering
\includegraphics[width=0.82\textwidth]{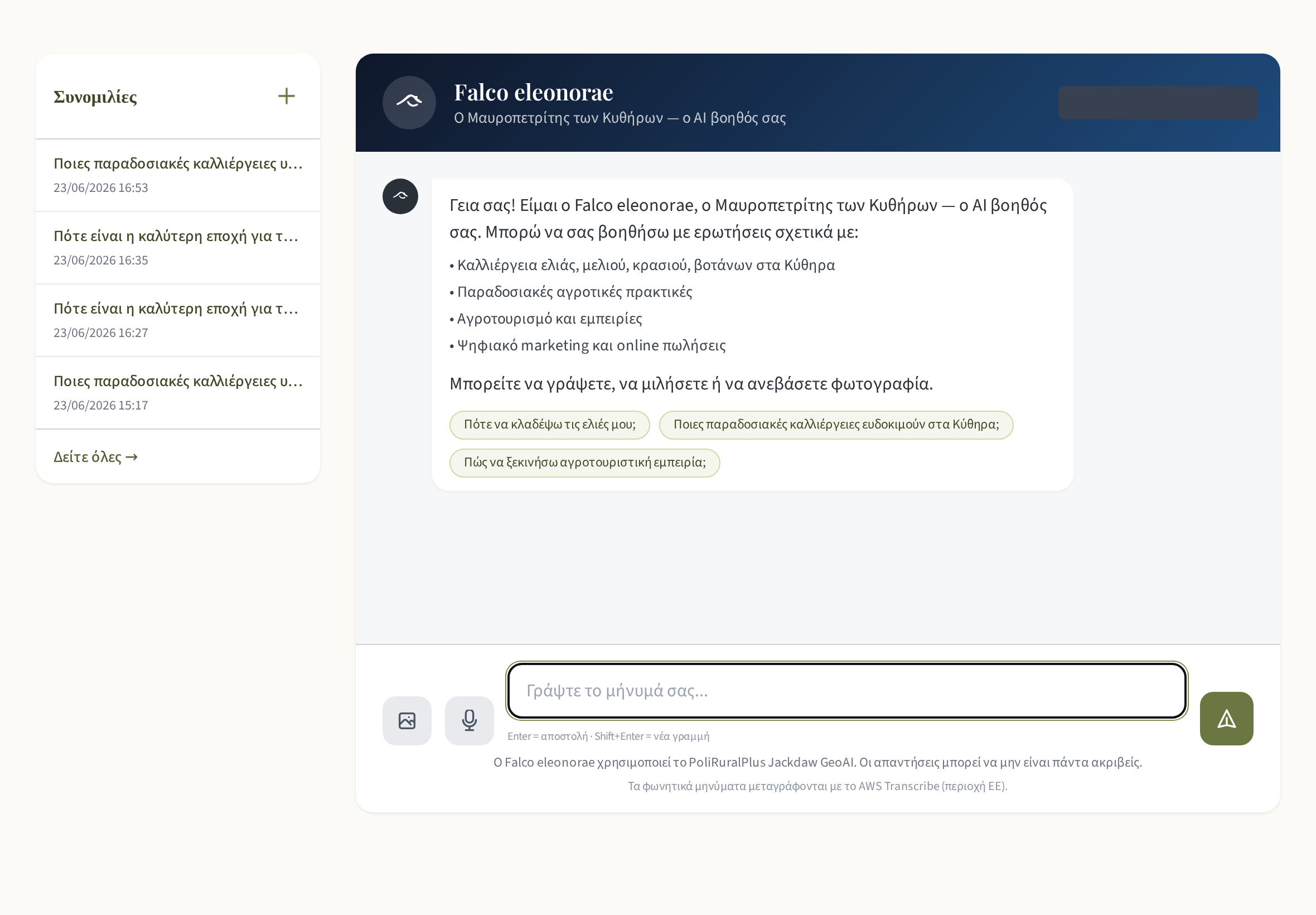}
\caption{The ``Falco eleonorae'' conversational assistant. The interface is Greek-primary, with an English toggle, a streamed-response area, a conversation-history sidebar, and microphone and image-attachment affordances beside the text box. (User account identifiers have been redacted.)}
\label{fig:chat_interface}
\end{figure}

The interface is where the design constraints of Section~\ref{sec:intro} are met in practice. It is a single-page chat experience embedded in the platform (Figure~\ref{fig:chat_interface}), built as a progressive web application so that it installs to a phone's home screen and degrades gracefully on poor connections.

\paragraph{Bilingual, Greek-primary interaction.} The entire interface is bilingual, with Greek as the primary language and English secondary; the active language also selects the speech-recognition language and a per-language clause in the system prompt, so that a Greek query is transcribed as Greek and answered in Greek. This is a direct response to the language constraint of Section~\ref{sec:intro} and to the AKIS literature's emphasis on meeting farmers in their own language~\citep{high2026aiextension}.

\paragraph{Streaming responses.} Answers are delivered over Server-Sent Events and rendered progressively, so the user sees the assistant ``typing'' rather than waiting for a complete reply, which matters when a tool-using turn takes appreciable time. Markdown in the answer is sanitised before rendering.

\begin{figure}[htbp]
\centering
\includegraphics[width=0.82\textwidth]{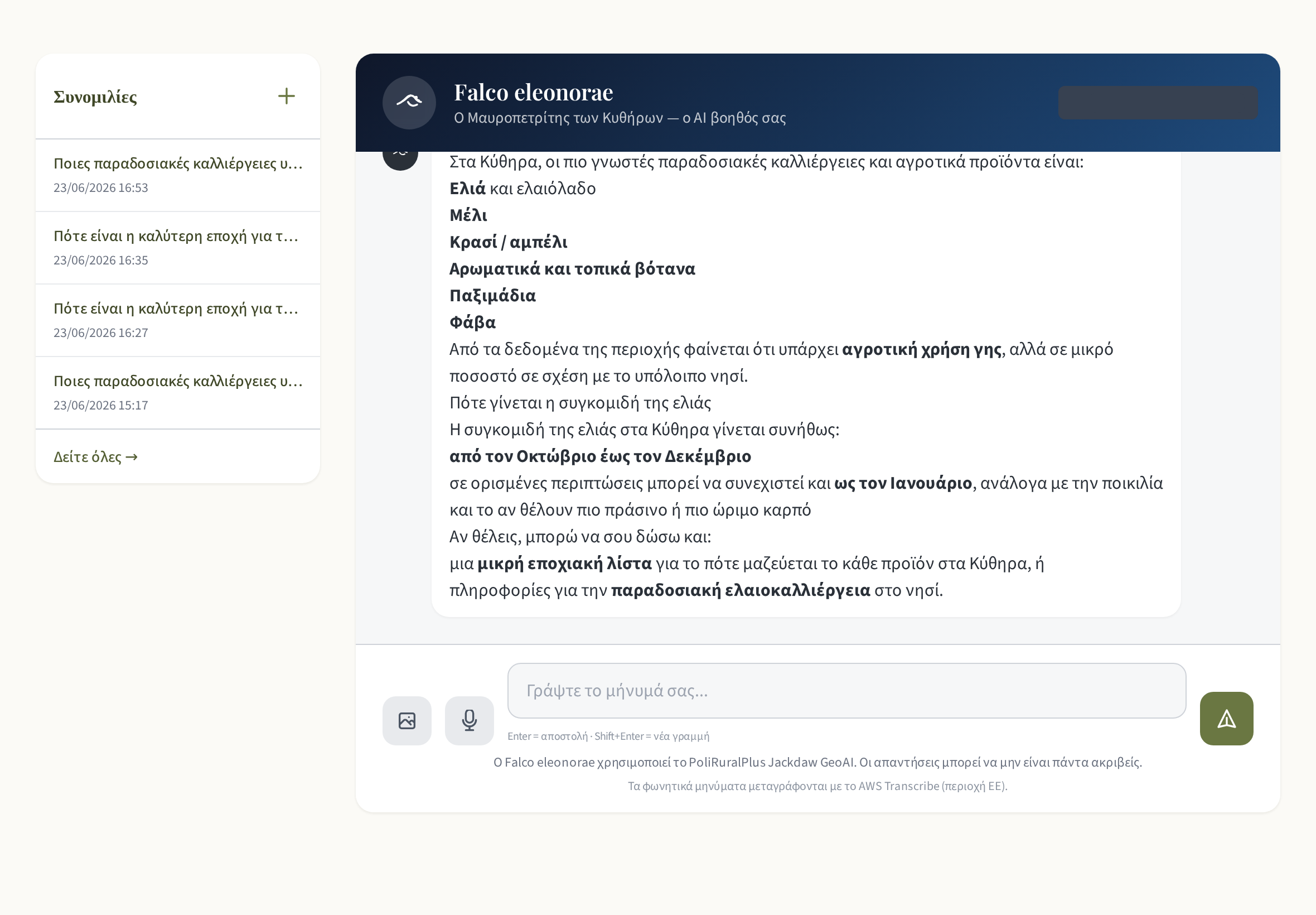}
\caption{A grounded exchange in the assistant: a Greek-language question about local crops and olive-harvest timing, answered with content anchored in the local data API via the \texttt{kythera} MCP tool. (User account identifiers have been redacted.)}
\label{fig:chat_conversation}
\end{figure}

\paragraph{Voice input.} A microphone control records a short utterance in the browser and submits it to an endpoint that streams it to the managed speech-to-text service; the returned transcript is inserted as an ordinary text turn (Figure~\ref{fig:chat_conversation}). Voice lowers the barrier for users uncomfortable with typing or with on-screen Greek keyboards, a population the digital-literacy constraint makes central.

\paragraph{Image input.} A user can attach a field photograph, for example of a leaf, a pest, or a soil condition. The platform passes the image to Claude Sonnet~4.5, which returns a short factual description; this description is folded into the user's next text turn as context, so that the upstream agent, which is itself text-only, can reason about the photograph. 

\paragraph{Conversation history.} Each user's conversations are retained and listed, and cross-turn continuity is preserved by passing an upstream thread identifier, so the agent retains context across turns within a conversation.

\section{Implementation, Security and Data Protection}
\label{sec:impl}

\paragraph{Proxy and streaming.} The assistant is implemented as a dedicated application within the platform. The streaming view holds an upstream connection open and relays SSE frames; the deployment's request timeout is set deliberately longer than the upstream read budget so that a worker is not killed mid-stream before the assistant turn is persisted. Errors that occur after streaming has begun are surfaced to the client as in-band SSE error events rather than as HTTP error codes.

\paragraph{Rate limiting.} The multimodal endpoints (voice transcription, image description) and the session-establishment endpoint are rate-limited per user, and the read-only data API is rate-limited per client, to bound cost and abuse.

\paragraph{Data protection.} Several measures address the trust-and-governance constraint. Speech transcription is pinned to an EU region, so audio and transcripts remain within the EU. The data API authenticates callers with a constant-time key comparison and fails closed in production. Editorial gating, with drafts and inactive records excluded and hidden prices suppressed, is enforced server-side, so the assistant cannot surface unpublished content or hidden prices. Chat content is included in the platform's account-deletion data-erasure routine. Inline scripts run under a content-security policy, and answer markdown and tool-supplied option text are sanitised to prevent injection from model or tool output.

\section{Discussion}
\label{sec:discussion}

\subsection{Applications for the pilot community}
\label{sec:apps}

The assistant supports a range of concrete tasks for the pilot community. These include seasonal guidance (``when do we harvest the olives?'') grounded in the local seasonal calendar; crop guidance on local varieties and practices; discovery of local products, agritourism experiences, and cooperatives, filterable to a map area; lookup of dialect and agronomic glossary terms; retrieval of training-lesson content; lightweight image-assisted observation of a crop condition; and voice-first interaction for users who prefer speaking to typing. In each case the value over a generic chatbot is the same, in that the answer is anchored in the island's own curated, editorially-controlled data, in the user's language, and within the correct geographic context.

\subsection{Relation to prior work}
\label{sec:relation}

Our grounding strategy is a concrete instance of tool-augmented retrieval~\citep{yao2023react} over MCP~\citep{mcp2025spec}, chosen over vector-store RAG~\citep{lewis2020rag} for the reasons set out in Section~\ref{sec:grounding}, since the knowledge is small, structured, curated, and frequently edited, a regime in which exact queries over a live API dominate approximate retrieval over a re-indexed embedding store. The design also operationalises the cautions of the agricultural-extension literature, with Tzachor et~al.'s insistence on grounding and human oversight~\citep{tzachor2023llm} and High et~al.'s finding that such tools work only when embedded in trusted, locally-controlled infrastructure~\citep{high2026aiextension} reflected, respectively, in the structured grounding and in keeping all local content under the pilot organisation's editorial control.

\subsection{Limitations and fragility}
\label{sec:limits}

We are explicit about the system's boundaries. \textbf{(i)~No formal evaluation yet.} This paper reports the system as built; we have not conducted a controlled user study, and the quantitative validation of answer quality, grounding faithfulness, latency, and user satisfaction with the pilot cohort is the principal item of future work. Informal integration testing against the live upstream service indicated that simple queries return in roughly fifteen to twenty seconds and multi-tool queries in roughly one minute, but these are anecdotal observations rather than measured results. \textbf{(ii)~Dependence on external services.} The conversational brain is upstream, so if it is unavailable, or the user has no valid upstream session, the assistant cannot answer, with no local fallback model. \textbf{(iii)~Upstream maturity.} During integration, several of the upstream agent's general geospatial services (for example climatology for the island coordinates) were observed to be unreliable, and the platform's own \texttt{kythera} MCP tool depends on an upstream deployment step to become live, so end-to-end grounding may not be fully active in production at all times. \textbf{(iv)~Stopgap authentication.} Establishing the upstream session by re-using the user's credentials, rather than a registered OAuth2 flow, is a deliberate but fragile interim measure that depends on the upstream login process and should be replaced by a registered client. \textbf{(v)~Single bounded vision use.} Image understanding is limited to short descriptions folded into text; it is not a diagnostic system, and should not be presented to farmers as one.

\section{Conclusions and Future Work}
\label{sec:conclusion}

We have described a conversational AI assistant embedded in a bilingual e-market platform, an LLM-powered, geospatially-grounded agronomic assistant for the smallholder farmers of Kythera and Antikythera. The assistant addresses the island-AKIS knowledge gap with three deliberate design choices: a thin Backend-for-Frontend architecture that hosts no model of its own and concentrates direct frontier-model use in a single bounded task; tool-augmented retrieval over a curated, read-only, bilingual data API, rather than vector-store RAG, as the grounding mechanism for small, structured, frequently-edited local knowledge; and a geospatial area-of-interest envelope that anchors every answer to the island. The interface is engineered for the realities of the pilot, being Greek-primary, multimodal (text, voice, image), streamed, and delivered as a progressive web app for poor connectivity, and the implementation is conservative about cost, data residency, and content governance.

The clear next step is a structured field evaluation with the pilot cohort, measuring grounding faithfulness, answer quality in Greek, latency, and user-reported usefulness, and feeding the results back into the system prompt, tool descriptions, and interface. Further work includes replacing the stopgap authentication with a registered OAuth2 client, hardening the dependence on upstream services, and expanding the curated local knowledge that the grounding tools expose. We offer the design as a transferable pattern for other remote and resource-constrained rural communities for whom a managed, grounded, multilingual assistant is more attainable, and more trustworthy, than a bespoke self-hosted model.

\section*{Acknowledgements}
The authors thank the \mbox{PoliRuralPlus} consortium for the Jackdaw agent infrastructure and the Kythera pilot partners for curating the local knowledge that grounds the assistant.

\section*{Funding}
This work was supported by the European Union's Horizon Europe research and innovation programme through the \mbox{PoliRuralPlus} project (grant agreement No.\ 101136910) via its DEVELOP cascade-funding open call. Views and opinions expressed are those of the authors only and do not necessarily reflect those of the European Union nor the granting authority.

\section*{Data and Software Availability}
The platform is deployed as a progressive web application, and the conversational assistant described here is available within its training section. The source code, the curated local data interface, and access to a demonstration instance are maintained by the project team and available on reasonable request, subject to the platform's data-protection obligations to the pilot community.

\bibliographystyle{unsrtnat}
\bibliography{references}

@article{tantaroudas2026interact,
  author    = {Tantaroudas, Nikolaos D. and McCracken, Andrew J. and Karachalios, Ilias and Papatheou, Evangelos},
  title     = {{INTERACT}: {AI}-powered extended reality platform for inclusive communication with real-time sign language translation and sentiment analysis},
  journal   = {Open Research Europe},
  volume    = {6},
  pages     = {71},
  year      = {2026},
  doi       = {10.12688/openreseurope.23201.1},
  note      = {[version 1; awaiting peer review]}
}

@misc{tantaroudas2026querying,
  author    = {Hontan, Valentin-Micu and Bunea, Andrei-Alexandru and Tantaroudas, Nikolaos Dimitrios and Popovici, Dan-Matei},
  title     = {Querying Structured Data Through Natural Language Using Language Models},
  year      = {2026},
  howpublished = {arXiv preprint arXiv:2604.03057},
  url       = {https://arxiv.org/abs/2604.03057}
}

@misc{tantaroudas2026career,
  author    = {Tantaroudas, Nikolaos D. and McCracken, Andrew J. and Karachalios, Ilias and Papatheou, Evangelos and Pastrikakis, Vasileios},
  title     = {{XR-CareerAssist}: An Immersive Platform for Personalised Career Guidance Leveraging Extended Reality and Multimodal {AI}},
  year      = {2026},
  howpublished = {arXiv preprint arXiv:2604.06901},
  url       = {https://arxiv.org/abs/2604.06901}
}

@inproceedings{tantaroudas2026sentinel,
  author    = {Tantaroudas, Nikolaos D. and Karachalios, Ilias and McCracken, Andrew},
  title     = {{SentinelSphere}: {AI}-Driven Cybersecurity Platform Combining Threat Detection with Security Awareness},
  booktitle = {Proceedings of the 21st International Conference on Cyber Warfare and Security (ICCWS 2026)},
  volume    = {21},
  pages     = {476--485},
  year      = {2026},
  doi       = {10.34190/iccws.21.1.4465}
}

@article{tzachor2023llm,
  author    = {Tzachor, Asaf and Devare, Medha and Richards, Catherine and Pypers, Pieter and Ghosh, Aniruddha and Koo, Jawoo and Johal, Sukhwinder and King, Brian},
  title     = {Large language models and agricultural extension services},
  journal   = {Nature Food},
  volume    = {4},
  pages     = {941--948},
  year      = {2023},
  doi       = {10.1038/s43016-023-00867-x}
}

@article{akis2024review,
  author    = {Kountios, Georgios and Kanakaris, Spyridon and Moulogianni, Christina and Bournaris, Thomas},
  title     = {Strengthening {AKIS} for Sustainable Agricultural Features: Insights and Innovations from the {European Union}: A Literature Review},
  journal   = {Sustainability},
  volume    = {16},
  number    = {16},
  pages     = {7068},
  year      = {2024},
  doi       = {10.3390/su16167068}
}

@article{high2026aiextension,
  author    = {High, Chris and Singh, Namita and Nemes, Guszt{\'a}v},
  title     = {Artificial Intelligence for Agricultural Extension: Supporting Transformative Learning Among Smallholder Farmers},
  journal   = {Journal of Development Policy and Practice},
  year      = {2026},
  doi       = {10.1177/24551333251345224}
}

@inproceedings{vaswani2017attention,
  author    = {Vaswani, Ashish and Shazeer, Noam and Parmar, Niki and Uszkoreit, Jakob and Jones, Llion and Gomez, Aidan N. and Kaiser, {\L}ukasz and Polosukhin, Illia},
  title     = {Attention Is All You Need},
  booktitle = {Advances in Neural Information Processing Systems (NeurIPS)},
  volume    = {30},
  year      = {2017},
  doi       = {10.48550/arXiv.1706.03762}
}

@article{brown2020language,
  author    = {Brown, Tom B. and Mann, Benjamin and Ryder, Nick and Subbiah, Melanie and others},
  title     = {Language Models are Few-Shot Learners},
  journal   = {Advances in Neural Information Processing Systems (NeurIPS)},
  volume    = {33},
  year      = {2020},
  doi       = {10.48550/arXiv.2005.14165}
}

@article{ouyang2022instructgpt,
  author    = {Ouyang, Long and Wu, Jeffrey and Jiang, Xu and Almeida, Diogo and Wainwright, Carroll L. and Mishkin, Pamela and others},
  title     = {Training language models to follow instructions with human feedback},
  journal   = {Advances in Neural Information Processing Systems (NeurIPS)},
  volume    = {35},
  year      = {2022},
  doi       = {10.48550/arXiv.2203.02155}
}

@article{bai2022constitutional,
  author    = {Bai, Yuntao and Kadavath, Saurav and Kundu, Sandipan and Askell, Amanda and Kernion, Jackson and others},
  title     = {Constitutional {AI}: Harmlessness from {AI} Feedback},
  journal   = {arXiv preprint arXiv:2212.08073},
  year      = {2022},
  doi       = {10.48550/arXiv.2212.08073}
}

@article{lewis2020rag,
  author    = {Lewis, Patrick and Perez, Ethan and Piktus, Aleksandra and Petroni, Fabio and Karpukhin, Vladimir and Goyal, Naman and K{\"u}ttler, Heinrich and Lewis, Mike and Yih, Wen-tau and Rockt{\"a}schel, Tim and Riedel, Sebastian and Kiela, Douwe},
  title     = {Retrieval-Augmented Generation for Knowledge-Intensive {NLP} Tasks},
  journal   = {Advances in Neural Information Processing Systems (NeurIPS)},
  volume    = {33},
  year      = {2020},
  doi       = {10.48550/arXiv.2005.11401}
}

@inproceedings{karpukhin2020dpr,
  author    = {Karpukhin, Vladimir and O{\u{g}}uz, Barlas and Min, Sewon and Lewis, Patrick and Wu, Ledell and Edunov, Sergey and Chen, Danqi and Yih, Wen-tau},
  title     = {Dense Passage Retrieval for Open-Domain Question Answering},
  booktitle = {Proceedings of the 2020 Conference on Empirical Methods in Natural Language Processing (EMNLP)},
  year      = {2020},
  doi       = {10.48550/arXiv.2004.04906}
}

@inproceedings{yao2023react,
  author    = {Yao, Shunyu and Zhao, Jeffrey and Yu, Dian and Du, Nan and Shafran, Izhak and Narasimhan, Karthik and Cao, Yuan},
  title     = {{ReAct}: Synergizing Reasoning and Acting in Language Models},
  booktitle = {International Conference on Learning Representations (ICLR)},
  year      = {2023},
  doi       = {10.48550/arXiv.2210.03629}
}

@techreport{openai2025gpt5card,
  author      = {{OpenAI}},
  title       = {{GPT-5} System Card},
  institution = {OpenAI},
  year        = {2025},
  month       = aug,
  url         = {https://cdn.openai.com/gpt-5-system-card.pdf}
}

@techreport{anthropic2025sonnet45,
  author      = {{Anthropic}},
  title       = {System Card: {Claude Sonnet 4.5}},
  institution = {Anthropic},
  year        = {2025},
  month       = sep,
  url         = {https://www.anthropic.com/claude-sonnet-4-5-system-card}
}

@misc{mcp2025spec,
  author       = {{Anthropic} and {Model Context Protocol Contributors}},
  title        = {Model Context Protocol Specification (revision 2025-11-25)},
  year         = {2025},
  howpublished = {\url{https://modelcontextprotocol.io/specification/2025-11-25}}
}

@misc{poliruralplus,
  author       = {{PoliRURAL Plus Consortium}},
  title        = {{PoliRURAL Plus}: Building Capacity for Foresight-Driven Rural Innovation (Horizon Europe)},
  year         = {2024},
  howpublished = {\url{https://poliruralplus.eu/}}
}

\end{document}